\documentclass{article}

\usepackage{amsthm}
\usepackage{amsmath}
\usepackage{ amssymb }
\usepackage{amssymb}
\usepackage{dsfont}
\usepackage{ bbold }
\usepackage{multirow}
\usepackage{pdfpages}
\usepackage{subcaption}
\usepackage{color}
\usepackage{verbatim}
\usepackage[round]{natbib} 
\usepackage[english]{babel}
\usepackage{float}
\usepackage[hidelinks]{hyperref}
\usepackage{rotating}
\usepackage{booktabs}
\usepackage{units}
\usepackage{mathrsfs}
\usepackage[margin=1.5in]{geometry}
\usepackage{adjustbox}
\usepackage{makecell}
\usepackage{mathrsfs}
\newcommand{\R}{\mathbf{R}}
\newcommand{\x}{\mathbf{x}}

\newcommand{\xx}{\mathbb{x}}
\newcommand{\dmat}{\mathbf{d}}
\newcommand{\dtilde}{\tilde{\mathbf{d}}}
\newcommand{\dtildep}{\tilde{ \dmat }_{\hspace{-.6mm}{}{/}}{}_{\scriptscriptstyle \hspace{-.6mm}\matp}}

\newcommand{\tht}{{\scriptscriptstyle{\textnormal{HT}}}}

\newcommand{\tgn}{{\scriptscriptstyle{\textnormal{N}}}}

\newcommand{\tgs}{{\scriptscriptstyle{\textnormal{GS}}}}
\newcommand{\tl}{{\scriptscriptstyle{\textnormal{T}}}}
\newcommand{\tm}{{\scriptscriptstyle{\textnormal{M}}}}

\newcommand{\twls}{{\scriptscriptstyle{\textnormal{WLS}}}}
\newcommand{\toc}{{\scriptscriptstyle{\textnormal{OC}}}}
\newcommand{\tb}{{\scriptscriptstyle{\textnormal{B}}}}

\newcommand{\matp}{\mathbf{p}}

\newcommand{\m}{\mathbf{m}}
\newcommand{\M}{\mathbf{M}}

\newcommand{\g}{\mathbf{g}}
\newcommand{\Omat}{\mathbf{O}}
\newcommand{\omat}{\mathbf{o}}

\newcommand{\W}{\mathbf{W}}

\newcommand{\w}{\mathbf{w}}

\newcommand{\I}{\mathbf{i}}
\newcommand{\E}{\text{\textnormal{E}}}
\newcommand{\V}{\text{\textnormal{V}}}
\newcommand{\diag}[1]{\textnormal{\footnotesize diag}\left(#1\right)}

\newcommand{\nNegOne}{n^{-1}}

\DeclareMathAlphabet{\mathpzc}{OT1}{pzc}{m}{it}

\newcommand{\oDIVp}{\mathbf{o}_{\scriptscriptstyle (0)\hspace{-.6mm}{}{/ \scriptscriptstyle \matp}}}

 \newcommand{\dsBDIVrootp}{\mathfrak{B}} 
  \newcommand{\dsbDIVrootp}{\mathfrak{b}} 

\newcommand{\bpi}{\boldsymbol{\pi}}

\newcommand{\bpiInv}{\bpi^{-1}}

\usepackage{array}
\newcolumntype{L}[1]{>{\arraybackslash}p{#1}}
\newcolumntype{C}[1]{>{\centering \arraybackslash}p{#1}}
\newcommand{\onesmat}{\mathds{1}}
\newcommand{\ones}[1]{ 1_{\scriptscriptstyle {#1}}}

\usepackage{etex,etoolbox}
\usepackage{amsthm,amssymb}
\usepackage{thmtools}
\usepackage{environ}
\usepackage{array}
\newcolumntype{H}{>{\setbox0=\hbox\bgroup}c<{\egroup}@{}}

\makeatletter
\providecommand{\@fourthoffour}[4]{#4}

\newcommand\fixstatement[2][\proofname\space of]{%
	\ifcsname thmt@original@#2\endcsname
	\AtEndEnvironment{#2}{%
		\xdef\pat@label{\expandafter\expandafter\expandafter
			\@fourthoffour\csname thmt@original@#2\endcsname\space\@currentlabel}%
		\xdef\pat@proofof{\@nameuse{pat@proofof@#2}}%
	}%
	\else
	\AtEndEnvironment{#2}{%
		\xdef\pat@label{\expandafter\expandafter\expandafter
			\@fourthoffour\csname #1\endcsname\space\@currentlabel}%
		\xdef\pat@proofof{\@nameuse{pat@proofof@#2}}%
	}%
	\fi
	\@namedef{pat@proofof@#2}{#1}%
}

\usepackage[affil-it]{authblk}

\declaretheorem[style=plain,name=Theorem, numberwithin=section]{theorem}

\newtheorem{corollary}{Corollary}[theorem]
\newtheorem{lemma}[theorem]{Lemma}

\newtheorem{remark}{Remark}
\newtheorem{definition}[theorem]{Definition}
\newtheorem{conjecture}[theorem]{Conjecture}
\fixstatement[Proof of]{theorem}
\fixstatement[Proof of]{lemma}
\usepackage{pgfplotstable}

\begin{document}

\title{Unifying Design-based Inference: 
			\\ A New Variance Estimation Principle
	\\	\small \vspace{5mm} WORKING PAPER 2 OF 4
}

\author{Joel A. Middleton{\footnote{
			Charles and Louise Travers Department of Political Science, \textit{University of California, Berkeley.} \\ \emph{{email:} joel.middleton@gmail.com}}} }

\maketitle

\pagebreak

\section{Introduction}
 
In Unifying paper 1 of 4 a generalized sandwich variance estimators was proposed that is applicable to virtually any experimental design, any linear estimator and any variance bound (defined therein), subsuming other well known sandwich variance estimators (which go by names such as ``robust", ``cluster robust'', ``heteroskedastic consistent", ``sandwich", ``White", ``Huber-White", ``HC", ``CR", etc.).  This paper presents two novel classes of variance estimators with superior properties, in the absence of parametric or semi-parametric assumptions. 

The first new class of estimator is the Oblo\v{z}en\`{e} Chleb\`{i}\v{z}ky (OC) variance estimators as a novel alternative to the generalized sandwich in Paper 1 of 4. That the OC concept is unlikely to arise from other, more standard, frameworks is manifestly true in light of the 40 year lacuna since White (1980). For any member of the generalized sandwich variance estimator class, there is an OC with the same expected value. 
The this alternative replaces a random matrix at the center with a nonrandom one. The second type of estimator is guaranteed conservative for the variance of the estimator and is based upon a similar principle of replacing a random matrix with its (nonrandom) expectation.

There is heuristic appeal in eliminating variance components from variance estimators, because it may mean that the OC may be more precise than the sandwich upon which it is based, though this may need not be true in general. In simulations based on a real randomized experiment, reductions in variance of variance estimators is substantial, and guaranteed conservative variance estimators do not suffer from small sample bias the way so-called robust, sandwich-type variance estimators do.

\section{Notation}


Consider a randomized experiment with $k$ treatment arms. The Neyman causal model (NCM) assumes that the units in the experimental study represent a finite population of size $n$. For a given outcome measure, call it $y$, each unit, $i$, responds with one of $k$ possible values in $\{y_{1i}, y_{2i}, ..., y_{ki} \}$, depending on their treatment assignment. The possible responses are referred to as the \textit{potential outcomes}. In the NCM these values are considered (nonrandom) constants, which stands in contrast to other, more common, formulations where potential outcomes are assumed to be sampled from some (possibly nonparametric) joint distribution. 

The only random elements in the NCM are the treatment assignment indicators  \linebreak $\{R_{1i}, R_{2i},...,R_{ki} \}$, and they determine which potential outcome will be observed by the researcher. Since a unit, $i$, can only be assigned to one arm of the experiment these random indicators must sum to unity, i.e., $R_{1i}+R_{2i}+...+R_{ki}=1$.
A standard representation of the \textit{observed} response for the $i^{th}$ unit under the NCM would be,
\begin{align*}
Y_i^{obs}= y_{1i}R_{1i}+y_{2i}R_{2i}+...+y_{ki}R_{ki},
\end{align*}
which is itself random, due to the assignment indicators. The observed data can then be represented as $\{Y^{obs}_i, R_{1i}, R_{2i},..., R_{ki}, x_i\}_{\forall i}$, where $x_i$ is an additional vector of $k$ covariates. Like the potential outcomes, $x_i$ is considered to be nonrandom, and, unlike the potential outcomes, does not depend on the assignment, which might be ensured, in practice, by collecting the covariates before the assignment.

Ideally, a researcher might like to know the difference between responses under various arms for a given individual, $i$

for example, $y_{2i}-y_{1i}$, or perhaps $\frac{1}{2}\left( y_{2i}-y_{1i} + y_{4i}-y_{3i} \right)$ and so on, which are examples of different \textit{contrasts} between potential outcomes. 

However, it is clear from the definition of $Y_i^{obs}$ that individual treatment effects are not identified since only one of the two potential outcomes can be observed. This is known as \textit{fundamental problem of causal inference} \citep{holland}. As a result, we often study averages of these values over the units of study. To continue with the above examples, a researcher might be satisfied to estimate averages over these individual treatment effects. Continuing the example, we have the average treatment effects $n^{-1} \sum_i  \left(y_{2i}-y_{1i}\right) $ and $ n^{-1} \sum_i \frac{1}{2} \left(y_{2i}-y_{1i} + y_{4i}-y_{3i} \right)$, respectively. 

To simplify notation, let $y_1$, $y_2$,...,$y_k$ represent length $n$ vectors of potential outcomes associated with each of the arms, with the $i^{th}$ element of each corresponding to the $i^{th}$ unit.  Next, define
\begin{align*}
y := \left(y_1' \hspace{2mm} y_2' \hspace{2mm} \hdots \hspace{2mm} y_k' \right)',
\end{align*}
which is the length $kn$, representing all $k$ potential outcomes for each of the $n$ units. 

Next, if we let $1_{\scriptscriptstyle n}$ be a $n$-length vector of ones, then a $kn \times k$ \textit{intercept matrix} can be defined as,
\begin{align*}
\onesmat := & \left[
\begin{matrix}
1_{\scriptscriptstyle n} &  & & 
\\  & 1_{\scriptscriptstyle n} &   & 
\\  &  & \ddots 
\\  &  & & 1_{\scriptscriptstyle n} 
\end{matrix} \right], 
\end{align*}
which, for example, allows us to express a $k$-length vector of means of the arms as as $\nNegOne \onesmat' y$ or, equivalently, $\left(\onesmat' \onesmat\right)^{-1} \onesmat' y$. 

Given a length-$k$ \textit{contrast vector}, $c$, we can contrast the means for the various arms.  In keeping with the running example with four assignment arms, if one is interested in the average difference in responses to the first two treatments, then $c=(-1\hspace{2mm} 1 \hspace{2mm} 0 \hspace{2mm} 0)'$ and we have $ \nNegOne c' \onesmat' y= n^{-1} \sum_i  \left(y_{2i}-y_{1i}\right) $. Likewise, if one chooses $c=(-\frac{1}{2}\hspace{2mm} \frac{1}{2} \hspace{2mm} -\frac{1}{2} \hspace{2mm} \frac{1}{2} )'$ then $\nNegOne c' \onesmat' y = n^{-1} \sum_i \frac{1}{2} \left(y_{2i}-y_{1i} + y_{4i}-y_{3i} \right)$. Sensible contrasts such as these meet the conditions $\sum_{j=1}^{k} c_j=0$ and $\sum_{j=1}^{k} | c_j | =2$, though this would not be strictly necessary, mathematically speaking.

Next define an $n \times n$ diagonal matrix that has all $n$ assignment indicators for treatment arm 1 on the diagonal, 
\begin{align*}
\R_1 :=&
\left[ \begin{matrix}
R_{11} \\ & R_{12} \\ & & \ddots \\& & &  R_{1i} \\ & & & & \ddots& \\ & & & & & R_{1n} 
\end{matrix}\right], \hspace{2mm}
\end{align*}
and define $\R_2$, $\R_3$, $\hdots$, $\R_k$ analogously. Arange these matrices to create the diagonal $kn \times kn$ matrix
\begin{align*}
\R :=&
\left[ \begin{matrix}
\R_{1} \\ & \R_{2} \\ & & \ddots \\& & &  \R_{k}
\end{matrix}\right] \hspace{2mm}
\end{align*}
and note the a $kn\times kn$ diagonal matrix of assignment probabilities can be written as $\bpi:=\E[\R]$,
with the first $n$ diagonal elements representing probabilities of assignment to arm 1, then the next $n$ diagonal elements are probabilities of assignment to arm 2 and so on. 

In this alternative notation the researcher can be said to observe the assignment, $\R$, the observed vector of outcomes, $\R y$, and also a matrix of $l$ pre-treatment covariates, $\x$, which has size $n \times l$. In a randomized experiment $\bpi$ is also observed (known) in many cases. When intractable analytically, however, it might be estimated to arbitrary precision by repeating the original randomization until a target level of precision is achieved.

For covariate adjusted estimators, it will also be convenient to define the $kn \times (k+l) $ matrix,
\begin{align*}
\xx := & \left[
\begin{matrix}
1_{\scriptscriptstyle n} & & & & \x
\\  & 1_{\scriptscriptstyle n} & & & \x
\\  &  & \ddots & & \vdots
\\  &  & & 1_{\scriptscriptstyle n} & \x
\end{matrix} \right].
\end{align*}

\begin{remark}
	For some cases, such adjusting for covariates separately by arm, it might be useful to define $\xx$ with $\x$ matrices arranged along a block-diagonal. In that case, it is prudent to stipulate that $\x$ have columns that sum to zero to avoid problems of coefficient interpretation \citep[cf.][]{lin, middleton18}. This will be discussed further in paper 3 of 4. 
\end{remark}

\section{The class of WLS estimators}\label{section.estimators}



This paper will discuss the class of estimators that can be characterized as weighted least squares (WLS), which includes the difference-of-means, OLS and Hajek estimators as special cases.  For the purposes of giving asymptotic variance expressions for this class, the Horvitz-Thompson estimator is also introduced. Paper 3 of 4 will introduce a wider class of estimators that will include these as special cases, and specifically examine their properties as point estimators. This paper is primarily concerned with a new approach to variance bound estimation, irrespective of what the estimator \textit{estimates}.

As in Paper 1 of 4, \textit{linear estimators} are defined as having the form,
\begin{align}\label{linear.est}
\widehat{\delta}_c := &  c' \W \R y,
\end{align}
where $\W$ a matrix with $kn$ columns and $k$ rows if it is an unadjusted esimator and $k+l$ rows if it is a covariate adjusted estimator. The length of the contrast vector, $c$, is equal to the number of rows in $\W$. The first $k$ entries of $c$ are the contrast values, followed by $l$ zeros in the case of a covariate adjusted estimator.

\begin{definition}[Horvitz-Thompson estimator]
The Horvitz-Thompson estimator written as in equation (\ref{linear.est}) with,
\begin{align*}
\W &=  \w^\tht :=\left(\onesmat'\onesmat\right)^{-1} \onesmat'\bpiInv .
\end{align*}
\end{definition}

\begin{definition}[WLS estimators]
WLS estimators can be written as in equation (\ref{linear.est}) with,
\begin{align*}
\W &= \W^\twls  := \left(\xx' \m \R \xx \right)^{-1}\xx' \m
\end{align*}
where the $kn \times kn$ matrix $\m$ must have nonegative entries on the diagonal (with some strictly positive) and zeros elsewhere.  Also, define
\begin{align*}
\w^{\twls} := & \left. \W^{\twls} \right|_{\R=\bpi}
\\ = & \left(\xx' \m \bpi \xx \right)^{-1}\xx' \m,
\end{align*}
which is $\W^{\twls}$ with $\bpi=\E\left[\R\right]$ replacing $\R$.
\end{definition}

\begin{remark}
	WLS gives OLS as a special case when $\m=\I_{kn}$ ($\I_{kn}$ is the identity matrix). If $\m=\I_{kn}$ and, in addition, $\xx=\onesmat$ (there are no covariates), WLS is equivalent to the difference-of-means.  If $\m=\bpiInv$ and $\xx=\onesmat$, then it is the Hajek estimator.  The covariate adjusted WLS with $\m=\bpiInv$ will be discussed further in paper 3 of 4, because it is algebraically equivalent to the generalized regression estimator introduced there. 
\end{remark}

\begin{theorem}\label{theorem.Taylor.is.HT}
	For a suitably chosen constant, $a$, and vector of constants, $z_c$, the first-order Taylor approximation for a WLS can be written as,
	\begin{align*}
	\widehat{\delta}^{\tl (\twls)}_c
	= & \hspace{1mm} a 
	+  n\ones{(k+l)}' \w^{\tht} \R  z_c,
	\end{align*}
	with 
	\begin{align*}
	z_c := \bpi \diag{u} {\w^\twls}' c ,
	\end{align*}
	and where
	\begin{align*}
	u := y- \xx b^\twls
	\end{align*}
	is the ``true" residual and
	\begin{align*}
	b^\twls := \w^{\twls} \bpi y
	\end{align*}
	is the ``true" WLS coefficient. 
\end{theorem}

\begin{remark}
	By ``true" it is meant to suggest that these are nonrandom quantities that, for large enough samples, represent approximately the centers of the distributions of observed residuals and WLS coefficient, respectively. 
\end{remark}

\begin{remark}
While the vectors $z_c$, $u$ and $b^\twls$ are not directly observable, the result is useful because it shows that the first-order Taylor approximation of a linear estimator is a Horvitz-Thompson estimator with contrast vector $c=n \ones{(k+l)}$ and ``outcome" vector $z_c$.  Expressing its variance will provide a basis for an asymptotically valid approximation. The constant $a$ is unimportant for the purposes of variance expressions, and so its explicit form is not given.
\end{remark}

\begin{corollary}[The Oblo\v{z}en\`{e} Chleb\`{i}\v{z}ky Principle]\label{corollary.z.insideout}
In Theorem \ref{theorem.Taylor.is.HT} we could have also given an equivalent expression for, $z_c$, as,
\begin{align*}
z_c^{equiv} := & \bpi \hspace{1mm} \diag{ c' \w^\twls } u 
\\ =& z_c.
\end{align*}
\end{corollary}

\begin{remark}
	Corollary \ref{corollary.z.insideout} gives the key insight in the development of the Oblo\v{z}en\`{e} Chleb\`{i}\v{z}ky variance estimators.  Residuals need not be in the center of the sandwich.
\end{remark}

\section{Variance bounds and their estimation}\label{section.var}

As in Paper 1 of 4, define $\ones{kn}$ as a $kn$-length vector of ones and the $kn \times kn$ ``design matrix" as,
\begin{align}\label{dmat}
\dmat:=\V \left( \ones{kn}' \bpiInv \R \right).
\end{align}
This is a variance-covariance matrix of weighted treatment assignments.

An exact expression for first-order Taylor approximations of linear estimators can be written as,
\begin{align} \label{var.general}
\V\left(\widehat{\delta}^{\tl (\twls)}_c \right) = z'_c \dmat z_c.
\end{align} 

\begin{definition}[Bounding Matrix]\label{def.varbound}
	Let $y$ be an outcome vector and let the contrast vector be $c=n \ones{k}$. Then the associated Horvitz-Thompson estimator is $n \ones{k} \w^{\tht}\R y$ with variance $y'\dmat y$. Then, the arbitrary $kn \times kn$ matrix, $\tilde{\dmat}$, is a ``bounding matrix" if, for all $y\in\mathds{R}^{kn}$, $y'\dmat y \leq  y'\tilde{\dmat}y$. 
\end{definition}

\begin{remark}
	For more on bounding matrices, see Paper 1 of 4. Unless otherwise noted, the Generalized Neyman Bounding matrix, $\dtilde=\dtilde^{\tgn}$, will be assumed.
\end{remark}

Now, define the $2n \times 2n$ matrix of probabilities and joint probabilities of assignment, 
\begin{align}
{\matp} := \E \left[\R 1_{\scriptscriptstyle 2n} 1'_{\scriptscriptstyle 2n} \R \right] \nonumber.
\end{align}
Next define an inverse probability weighted version of bounding matrix, $\tilde{ \dmat }$, as
\begin{align} 
\dtildep  := \tilde{ \dmat } / \matp
\end{align}
with $/$ denoting element-wise division defined such that division by zero equals zero.  

Hypothetically, if the observed outcome vector was $\R z_c$, then the Horvitz-Thompson estimator with $c=n \ones{k}$ has variance $z_c' \tilde{ \dmat }z_c$, which could be estimated unbiasedly by
\begin{align}\label{var_lin_est}
\widehat{\tilde{\V}}\left(\widehat{\delta}^{\tht }_{n1_k} \right) :=  z_c' \R \dtildep \R z_c.
\end{align}
Of course, the vector $z_c$ is defined by quantities which are not directly observed. However, we might identify random quantities that approximate $z_c$ and $u$ as
\begin{align*}
Z_c := & \bpi \diag{U} {\W^\twls}' c 
\\ = & \bpi \diag{c'{\W^\twls} } U
\end{align*}
and
\begin{align*}
U := & y- \xx \W^\twls \R y,
\end{align*}
respectively. 

\begin{definition}[The Generalized Sandwich Variance Estimator]
The Generalized Sandwich Variance Estimator is,
\begin{align}\label{var_sand_est}
\widehat{\tilde{\V}}{}^{\tgs}\left(\widehat{\delta}^{\tl(\twls) }_c \right) := & Z_c'\R \dtildep \R Z_c
\\=& c' \W^\twls \hspace{1mm} \diag{\R U} \left(\bpi \dtildep \bpi \right) \diag{\R U} {\W^\twls}' c \nonumber
\\ = &  y' \bigg  \{ \M \hspace{1mm} \diag{c' \W^\twls}\left(\bpi \dtildep \bpi \right) \hspace{1mm} \diag{c' \W^\twls}  \M \bigg \} y \nonumber 
\\ = & y' \Omat_{\scriptscriptstyle (0)} y \nonumber 
\end{align}
where $\M :=(\R- \R \xx \W^\twls \R )$ is a ``residual maker" matrix and $\Omat_{\scriptscriptstyle (0)} $ denotes the random matrix in curly brackets in the line just above with the subscript $(0)$ differentiating $\Omat_{\scriptscriptstyle (0)}$ from $\Omat_{\scriptscriptstyle (1)}$ and $\Omat_{\scriptscriptstyle (2)}$, introduced below. 
\end{definition}
The third line uses the Oblo\v{z}en\`{e} Chleb\`{i}\v{z}ky principle from Corollary \ref{corollary.z.insideout} and the equality $\R U=\M y$.

\begin{remark}
	The generalized sandwich is equivalent to Eicker-Huber-White variance estimator under Bernoulli designs and $\m=\I_{kn}$. Likewise it is algebraically equivalent to ``cluster robust" standard errors (CR0) under Bernoulli assignment of clusters and $\m=\I_{kn}$ (see Paper 1 of 4 for more details). Refinements for degrees-of-freedom and leverage are easily accommodated, and simulations will compare sandwiches HC0, HC1, and HC2 to the parallel OC estimators.
\end{remark}

\section{Oblo\v{z}en\`{e} Chleb\`{i}\v{z}ky variance bound estimators}

The last line of equation (\ref{var_sand_est}) has used the Oblo\v{z}en\`{e} Chleb\`{i}\v{z}ky principle from Corrolary \ref{corollary.z.insideout} to rearrange terms in the generalized sandwich. Written like this, it is easy to see that the expectation of the generalized sandwich is,
\begin{align} \label{var_gs_exp}
\E \left[ \widehat{\tilde{\V}}{}^{\tgs}\left(\widehat{\delta}^{\tl(\twls) }_c \right) \right] 
= & y' \E\left[\Omat_{\scriptscriptstyle (0)} \right] y \nonumber 
\\ = & y' \omat_{\scriptscriptstyle (0)} y 
\end{align}
with $\omat_{\scriptscriptstyle (0)}:=\E \left[ \Omat_{\scriptscriptstyle (0)} \right]$. The matrix $\omat_{\scriptscriptstyle (0)}$ can be  computed for small samples. For larger samples, it may be simulated to arbitrary precision by drawing from the randomization distribution.

\subsection{OC0: A ``baseline'' variance estimator that is not invariant}

\begin{definition}[OC0]
	Next, an estimator with the same expected value as the Generalized Sandwich in Equation \textnormal{(\ref{var_sand_est})}, is the OC0 estimator,
\begin{align*}
\widehat{\tilde{\V}}^{\toc 0}\left(\widehat{\delta}^{\tl(\twls) }_c\right) :=& y'\R \oDIVp{}  \R y, 
\end{align*}
with $\oDIVp := \omat_{\scriptscriptstyle (0)} / \matp$ where, as above, ``$/$" is element-wise division with division by zero resolving to zero.  
\end{definition}
The key insight with respect to OC0 is that it is an unbiased estimator of the expected value of the Generalized Sandwich, i.e.,
\begin{align*}
\E \left[\widehat{\tilde{\V}}^{\toc 0}\left(\widehat{\delta}^{\tl(\twls) }_c\right)\right] =& \E \left[ \widehat{\tilde{\V}}{}^{\tgs}\left(\widehat{\delta}^{\tl(\twls) }_c \right) \right],
\end{align*}
because $\E \left[\R \oDIVp \R \right]= \omat_{\scriptscriptstyle (0)}$.

OC0 is a Horvitz-Thompson estimator, and therefore lacks invariance to location shifts in the outcome variable. Likewise, its variance may depend on the scaling of $y$ and so it may also be less precise than covariate-adjusted alternatives. In the next sections, refinements that are invariant are considered.  

\subsection{OC1: Covariate-adjustment achieves invariance, but introduces bias}

To improve the precision of OC0, consider the covariate adjusted alternative,
\begin{align}\label{OC1}
\widehat{\tilde{\V}}^{\toc \scriptscriptstyle 1 }\left(\widehat{\delta}^{\tl(\twls) }_c\right) 
: =& y' \big \{ \M \oDIVp \M \big  \} y
\\ =& y' \mathbf{O}_{\scriptscriptstyle (1)} y
\end{align}
with $\M :=(\R- \R \xx \W^\twls \R )$, as above, the residual maker for WLS. Due to the covariate adjustment, OC1 may be more precise than OC0, and OC1 is also invariant to location shifts in $y$.  It is, nonetheless, biased for the mean of the Generalized Sandwich. In the next section, a bias correction is considered.

\subsection{OC2: An unbiased and invariant OC variance estimator}
Next the bias of OC1 relative to the expected value of the Generalized Sandwich variance,
\begin{align}\label{bias0}
\mathbf{Bias}^{\left(\tgs\right)} \left[\widehat{\tilde{\V}}^{\toc \scriptscriptstyle 1 }
\left(\widehat{\delta}^{\tl(\twls) }_c\right)\right]
=& \E \left[ \widehat{\tilde{\V}}^{\toc \scriptscriptstyle 1 }\left(\widehat{\delta}^{\tl(\twls) }_c\right)  - \widehat{\tilde{\V}}^{\tgs }\left(\widehat{\delta}^{\tl(\twls) }_c\right) \right] \nonumber
\\=& y' \bigg \{ \E  \left[ \M \oDIVp \M    - \R \oDIVp \R \right] \bigg \} y .
\end{align}

Next, define the degree 4, $kn \times kn \times kn \times kn$ tensor,
\begin{align*}
\dsBDIVrootp^{abcd} := \left( \M^{ab} \cdot \M^{cd} - \R^{ab} \cdot \R^{cd} \right) \circ \big( {\matp^{ad}} \cdot {\matp^{bc}} \big)^{\circ \tiny{-}\frac{1}{2}}, 
\end{align*}
with $a, b, c$ and $d$ giving labels for dimensions, $\cdot$ represents the tensor multiplication using Einstein's convention of inner product for dimensions with identical labels and outer product otherwise, $\circ$ is element-wise (Hadamard) multiplication and $(.)^{\circ \tiny{-}\frac{1}{2}}$ is element wise exponentiation by $-\frac{1}{2}$, with division by zero resolving to zero.  

With this tensor notation, the bias in Equation (\ref{bias0}) can be written,
\begin{align*}
\mathbf{Bias}^{\left(\tgs\right)} \left[\widehat{\tilde{\V}}^{\toc \scriptscriptstyle 0. \tb }
\left(\widehat{\delta}^{\tl(\twls) }_c\right)\right]
=& y' \bigg \{ \left( \E \left[ \dsBDIVrootp^{abcd} \right] \circ \left({\matp^{ad} \cdot \matp^{bc}}\right)^{\circ \frac{1}{2}} \right) \hspace{1mm}\cdot \hspace{1mm}  \oDIVp^{bc} \bigg \}  y
\\ =& y' \bigg \{ \left( \dsbDIVrootp^{abcd} \circ \left({\matp^{ad} \cdot \matp^{bc}}\right)^{\circ \frac{1}{2}} \right) \hspace{1mm}\cdot \hspace{1mm} \oDIVp^{bc} \bigg \}  y
\end{align*}
with $\dsbDIVrootp^{abcd}:= \E \left[ \dsBDIVrootp^{abcd} \right]$, which can be estimated to arbitrary precision using repeated randomizations. 

Next taking the tensor SVD we can decompose as follows,
\begin{align*}
\dsbDIVrootp^{abcd} =& - \mathfrak{u}^{ade} \cdot \lambda^e \cdot {\mathfrak{u}^{bce}} 
\\ =& - \mathfrak{u}^{ade}_{\scriptscriptstyle (0<\lambda<1)} \cdot \lambda^e_{\scriptscriptstyle (0<\lambda<1)} \cdot \mathfrak{u}^{bce}_{\scriptscriptstyle (0<\lambda<1)} 
\hspace{2mm} - \mathfrak{u}^{ade}_{\scriptscriptstyle (\lambda \geq 1)} \cdot \lambda^e_{\scriptscriptstyle (\lambda \geq 1)} \cdot \mathfrak{u}^{bce}_{\scriptscriptstyle (\lambda \geq 1)}
\\ = & \hspace{2mm}\dsbDIVrootp^{abcd}_{\scriptscriptstyle (0<\lambda<1)} \hspace{2mm} + \hspace{2mm} \dsbDIVrootp^{abcd}_{\scriptscriptstyle (\lambda \geq 1)},
\end{align*}
with $\lambda^e$ representing a length $k^2 n^2$ vector (tensor of degree 1) of singular values and $\mathfrak{u}^{ade}$ as a $kn \times kn \times k^2n^2$ tensor. In the second line, the tensor is decomposed into two terms. The first term, with subscripts $\left(0<\lambda <1\right)$, is the tensor constructed from the slices of $\mathfrak{u}$ corresponding to singular values greater than 0 but less than 1. The second term, with subscripts $\left(\lambda \geq 1\right)$, is constructed from the slices of $\mathfrak{u}$ corresponding to singular values greater than or equal to 1.  Note that the orthogonality property, such that the tensor product ${\mathfrak{u}^{ade}} \cdot \mathfrak{u}^{adf}$ results in the identity matrix, $\I^{ef}$.

\begin{conjecture}
	The singular values, $\lambda^e$, are bounded by zero and one, i.e., $0 \leq \lambda^e \leq 1$. 
\end{conjecture}

\begin{lemma}
	The infinite tensor series
	\begin{align*}
	{\dsbDIVrootp^{abcd}_{(\infty)}} \hspace{2mm} := & \hspace{2mm}
	\dsbDIVrootp^{abcd}_{\scriptscriptstyle (0<\lambda<1)} \hspace{2mm}- \hspace{2mm} \dsbDIVrootp^{aefd}_{\scriptscriptstyle (0<\lambda<1)}\cdot \dsbDIVrootp^{ebcf}_{\scriptscriptstyle (0<\lambda<1)} 
	\hspace{2mm}+ \hspace{2mm} \dsbDIVrootp^{aefd}_{\scriptscriptstyle (0<\lambda<1)}\cdot \dsbDIVrootp^{eghf}_{\scriptscriptstyle (0<\lambda<1)} \cdot 
	\dsbDIVrootp^{gbch}_{\scriptscriptstyle (0<\lambda<1)} \hspace{2mm}
	\\ & \hspace{4mm} - \hspace{2mm} \dsbDIVrootp^{aefd}_{\scriptscriptstyle (0<\lambda<1)}\cdot \dsbDIVrootp^{eghf}_{\scriptscriptstyle (0<\lambda<1)} \cdot \dsbDIVrootp^{gijh}_{\scriptscriptstyle (0<\lambda<1)} \cdot \dsbDIVrootp^{ibcj}_{\scriptscriptstyle (0<\lambda<1)} 
	\hspace{2mm}+ \hspace{2mm}
	\dots 
	\end{align*}
	has closed form,
	\begin{align*}
	{\dsbDIVrootp^{abcd}_{ (\infty)}} =
	- \mathfrak{u}^{ade}_{\scriptscriptstyle (0<\lambda<1)} \cdot \phi ^e_{\scriptscriptstyle (0<\lambda<1)} \cdot \mathfrak{u}^{bce}_{\scriptscriptstyle (0<\lambda<1)} 
	\hspace{2mm} 
	\end{align*}
	where $\phi ^e_{\scriptscriptstyle (0<\lambda<1)} := \lambda^e_{\scriptscriptstyle (0<\lambda<1)} / (\ones{} - \lambda^e_{\scriptscriptstyle (0<\lambda<1)})$ with $/$ representing element-wise division.
\end{lemma}

\begin{theorem}
	An unbiased estimator of the bias of OC1 as an estimator of the mean of the Generalized Sandwich is,
	\begin{align}\label{bias.est}
	\widehat{\mathbf{Bias}}^{\left(\tgs\right)} \left[\widehat{\tilde{\V}}^{\toc \scriptscriptstyle 1 }
	\left(\widehat{\delta}^{\tl(\twls) }_c\right)\right]
	:=& y' \M \bigg \{ \left( \dsbDIVrootp^{abcd}_{(\infty)} \circ \left({\matp^{ad}}^{\circ -\frac{1}{2}} \cdot {\matp^{bc}}^{\circ \frac{1}{2}} \right)  \right) \hspace{1mm}\cdot \hspace{1mm} \oDIVp^{bc} \bigg \} \M y 
	\\ & + y' \R \bigg \{ \left(  \dsbDIVrootp^{abcd}_{(\lambda \geq 1)} \circ \left({\matp^{ad}}^{\circ -\frac{1}{2}} \cdot {\matp^{bc}}^{\circ \frac{1}{2}} \right)  \right) \hspace{1mm}\cdot \hspace{1mm} \oDIVp^{bc} \bigg \}\R y ,\nonumber
	\end{align}
where $\M$ is the residual-maker, as above. 
\end{theorem}

\begin{conjecture}\label{conjecture.zeroInvariantTerm}
	The second term in Equation \textnormal{(\ref{bias.est})} is zero for all randomizations.  Hence, 
	\begin{align*}\label{bias.est.oneterm}
	\widehat{\mathbf{Bias}}^{\left(\tgs\right)} \left[\widehat{\tilde{\V}}^{\toc \scriptscriptstyle 1 }
	\left(\widehat{\delta}^{\tl(\twls) }_c\right)\right]
	=& y' \M \bigg \{ \left( \dsbDIVrootp^{abcd}_{(\infty)} \circ \left({\matp^{ad}}^{\circ -\frac{1}{2}} \cdot {\matp^{bc}}^{\circ \frac{1}{2}} \right)  \right) \hspace{1mm}\cdot \hspace{1mm} \oDIVp^{bc} \bigg \} \M y 
	\end{align*}
	is an unbiased estimator of the bias \textnormal{OC1} relative to the Generalized Sandwich.
\end{conjecture}

\begin{definition}[OC2]
	The OC2 variance estimator is, 
	\begin{align*}
	\widehat{\tilde{\V}}^{\toc \scriptscriptstyle 2}
	\left(\widehat{\delta}^{\tl(\twls) }_c\right)
	:=& y' \M \oDIVp \M y - 
	y' \M \bigg \{ \left( \dsbDIVrootp^{abcd}_{(\infty)} \circ \left( {\matp^{ad}}^{\circ -\frac{1}{2}} \cdot {\matp^{bc}}^{\circ \frac{1}{2}} 
	  \right) \right) \hspace{1mm}\cdot \hspace{1mm} \oDIVp^{bc} \bigg \}\M  y
	  \\	=& 
	  y' \Bigg \{ \M  \Bigg ( \bigg ( \I_{kn}^{ab} \cdot \I_{kn}^{cd} - \dsbDIVrootp^{abcd}_{(\infty)} \circ \left( {\matp^{ad}}^{\circ -\frac{1}{2}} \cdot {\matp^{bc}}^{\circ \frac{1}{2}} 
	  \right) \bigg )  \hspace{1mm}\cdot \hspace{1mm} \oDIVp^{bc} \Bigg ) \M \Bigg  \} y 
	  \\ =& y' \mathbf{O}_{\scriptscriptstyle (2)} y
	\end{align*}
\end{definition}

\begin{remark}
	If true, Conjecture \textnormal{(\ref{conjecture.zeroInvariantTerm})} implies that OC2 is unbiased as an estimator of the Generalized Sandwich variance.  The presence of the residual-maker matrix, $\M$, means that OC2 is invariant to location shifts in the outcome, $y$.
\end{remark}

\subsection{Comparing the precision of OC2 and GS}
Using tensor notation, the variance of the OC2 variance estimator can be written, 
	\begin{align*}
	\V\left(\widehat{\tilde{\V}}^{\toc \scriptscriptstyle 2}\right)
	=&  y^a \cdot y^b \cdot \bigg \{ \E \left[ \mathbf{O}^{ab}_{\scriptscriptstyle (2)} \cdot \mathbf{O}^{cd}_{\scriptscriptstyle (2)} \right] - \E \left[ \mathbf{O}^{ab}_{\scriptscriptstyle (2)} \right]\cdot \E \left[ \mathbf{O}^{cd}_{\scriptscriptstyle (2)} \right] \bigg \}  \cdot y^c \cdot y^d.
	\end{align*}
Similarly, the variance of the GS variance estimator can be written,
\begin{align*}
\V\left(\widehat{\tilde{\V}}^{\tgs \scriptscriptstyle }\right)
=&  y^a \cdot y^b \cdot \bigg \{ \E \left[ \mathbf{O}^{ab}_{\scriptscriptstyle (0)} \cdot \mathbf{O}^{cd}_{\scriptscriptstyle (0)} \right] - \E \left[ \mathbf{O}^{ab}_{\scriptscriptstyle (0)} \right]\cdot \E \left[ \mathbf{O}^{cd}_{\scriptscriptstyle (0)} \right] \bigg \} \cdot y^c \cdot y^d.
\end{align*}
So, with Conjecture (\ref{conjecture.zeroInvariantTerm}), we have the difference of variances, 
\begin{align*}
	\V\left(\widehat{\tilde{\V}}^{\tgs \scriptscriptstyle }\right)-\V\left(\widehat{\tilde{\V}}^{\toc \scriptscriptstyle 2}\right) 
	= y^a \cdot y^b \cdot \bigg \{ \E \left[ \mathbf{O}^{ab}_{\scriptscriptstyle (0)}\cdot \mathbf{O}^{cd}_{\scriptscriptstyle (0)}\right]  -\E \left[\mathbf{O}^{ab}_{\scriptscriptstyle (2)} \cdot \mathbf{O}^{cd}_{\scriptscriptstyle (2sim)} \right] \bigg \} \cdot y^c \cdot y^d.
\end{align*}
with $\E \left[ \mathbf{O}^{ab}_{\scriptscriptstyle (0)}\cdot \mathbf{O}^{cd}_{\scriptscriptstyle (0)}\right]$ and $\E \left[\mathbf{O}^{ab}_{\scriptscriptstyle (2)} \cdot \mathbf{O}^{cd}_{\scriptscriptstyle (2sim)} \right] $ computable in small samples or to arbitrary precision through repeated randomizations. 

\section{Guaranteed conservative variance estimator}\label{section.GC}

From first principles, the exact variance of a linear estimators of the form given in equation (\ref{linear.est}) is,
\begin{align*}
\V \left( \widehat{\delta_c} \right) = & \E \left[ y'\R \W' c c' \W \R y \right]-\E \left[ y'\R \W' c  \right] \E \left[c' \W \R y \right]
\\ = & y' \bigg \{ \E \left[ \R \W' c c' \W \R \right]-\E \left[ \R \W' c  \right] \E \left[c' \W \R \right] \bigg \} y
\\ = & y' \g y
\end{align*}
where $\g$ is defined as the matrix inside the curly brackets in the line above. The matrix, $\g$, may be computed exactly for small studies or to arbitrary precision by repeating the randomization until desired precision is achieved.

As before, some terms in this quadratic are not observed, and a suitable bound can be found generalizing one of the methods from Paper 1 of 4 (Neyman, Aronow-Samii or Middleton). Algorithm 4.7 from Paper 1 of 4 will be slightly modified and applied to the simulations below.

With Conjecture 

\section{Data from Paluck and Green (2009) }

\cite{paluckgreen} pair-randomized 14 villages (7 pairs) in post-genocide Rwanda to receive one of two possible radio programs. Half of the villages were exposed to a program aimed at ``discouraging blind obedience and reliance on direction from authorities and promoting independent thought and collective action in problem solving" (treatment). The other half were exposed to a radio program about health (control).

The analysis of the properties of the OC and GS variance estimators are compared in the spirit of (permutation/rerandomization-based) simulation. The analysis is simulation in the sense that it makes untestable assumptions about the outcomes that would have been observed under alternative assignments (e.g., the values of missing potential outcomes). The appeal of using real data, rather than generating data from a contrived DGP, however, is that it may provide a more realistic data set and the analysis can reflect the design actually used.

\subsection{Randomization method}

The original paired-cluster design was simulated and the estimator is OLS as in the original paper. With 7 pairs this results in $2^7=128$ possible randomizations.

\subsection{Data}

Analysis used the same 497 cases analyzed in model 1 of Table 4 in \cite{paluckgreen}. Number of units by village pair and assignment arm are given in Table \ref{table.clusterSizes}. The average cluster size was 35.5. The minimum and maximum cluster sizes were 20 and 43, respectively.

Five cases had missing values for \textit{age}, which were mean-imputed for this analysis so that the same units to be used in both specifications. By contrast, the original analysis dropped these five cases when \textit{age }was included in the OLS.

\begin{table}[ht]
	\caption{Number of units by village pair and original assignment}
	\centering 
	\begin{tabular}{rrrrrrrr}\label{table.clusterSizes}
		& \multicolumn{7}{c}{Village Pair}  \\
	\cline{2-8}	& 1 & 2 & 3 & 4 & 5 & 6 & 7 \\ 
		\hline
		Control &  37 &  39 &  39 &  39 &  33 &  37 &  43 \\ 
		Treatment &  33 &  37 &  36 &  37 &  38 &  20 &  29 \\ 
		\hline
	\end{tabular}
\end{table}

\subsection{Outcome measure}

The outcome is a post-treatment measure of social distance. The original effect estimates using OLS were presented in Table 4 of \cite{paluckgreen}. The social distance measure was an index created by combining four survey items, each of which could take on integer values of 1,2,3 or 4. The resulting measure, an average of the items, also ranged from 1 to 4, but could take on 13 possible values.

Additionally the outcome, y, was location shifted by subtracting the mid-point on the scale, (max($y$)-min($y$))/2, such that the scale ranged from -1.5 to 1.5 rather than 1 to 4. One motivation for this is that non-invariant estimators such as OC$J$ and GC$J$ can, themselves, have high variance when the mean of $y$ is shifted away from zero.  

Another motivation is that when scale is bounded, and thus max($y$) and min($y$) are known, shifting by the scale midpoint introduces an invariance property.  For example, subtracting off the scale midpoint results in the same observed values whether a researcher records the original outcomes on a scale of integers $\{0,1,2,3\}$ or $\{1, 2, 3,4\}$. 

Another option might be to subtract the observed sample mean from each outcome.  However, this leads to a shift that may depend on the randomization, when there are treatment effects, and so the practice can introduce biases. 

\subsection{Variance estimators compared}

The Generalized Neyman bounding matrix, given in section 4.1 of Paper 1 of 4 \citep{middleton20_1of4}, $\dtilde^\tgn$, is not applicable in this case because of unobservable cells in main block diagonal of the design matrix, $\dmat$.  Therefore simulations use the bounding matrix, $\dtilde^\tm$, which is based on Algorithm 4.7 in Paper 1 of 4.  It provides a tighter bound than the Aronow-Samii bounding matrix because clusters were pair-randomized, a result that is similar to the one given for pair-randomization of units in the example provided in section 4.4 of that paper.

The variance estimators are based upon the Generalized Sandwich (GS) given in equation (\ref{var_sand_est}) in Section \ref{section.var}, the Oblo\v{z}en\`{e} Chleb\`{i}\v{z}ky-based, given in Section \ref{subsection.OCJ} ($J=10$), and the Guaranteed Conservative, given in Section \ref{section.GC} ($J=10$).

\subsection{Covariate specifications}

The OC and GS variance estimators were examined with and without covariates. The ``no covariates" OLS included only intercepts, so that $\xx=\onesmat$. This is equivalently the difference-of-means. The second specification included the additional covariates in column 3 of Table 4 in the original: \textit{displaced by violence}, \textit{sex}, \textit{age}, and \textit{radio lisening habits}. 

For both of these specifications, classic Cluster Robust (CR) variance estimators also included fixed-effects for cluster-pairs as in columns 2 and 3 of Table 4 of \cite{paluckgreen}. The typical motivation for including the fixed effects for blocks/pairs is to obtain standard errors that reflect this design feature.  These are unnecessary in the case of OC and GC variance estimators because the paired-cluster design is already accounted for in the design matrix, $\dmat$.

\subsection{Simulation assumptions}

The assumptions of NCM apply, for example, assuming that potential outcomes are fixed (non-stochastic) and depend only on assignment. Also the sharp null, $y_{0i}=y_{1i}$ for all $i$, is invoked in order to impute missing potential outcomes.\footnote{Simulating under the sharp null seems reasonable for two reasons. First, the estimated average effects (-.029 to -.041, depending on specification) are small relative to the range of the measure (1.0\% to 1.4\%, respectively) and also to its standard deviation (3.6\% to 5.1\%, respectively). Second, and more importantly, simulating under the sharp null (or constant treatment effects more generally) is useful because the variance and its (identified) bound will be equal. The implication is that any \textit{downward} bias of the variance estimator that is attributable to borrowing the variance of the first-order approximation, and/or to using a ``plug-in'' estimator thereof, will be evident in simulation. By contrast, downward bias can be masked (i.e., offset) under heterogeneous effects because the \textit{bound} will tend to be conservative in that case.  So, simulating under the sharp null gives a ``worst-case" assessment of anti-conservative bias.} The simulation also proceeds as if the survey response is not stochastic and whether a unit participates in the survey is not dependent on treatment assignments.\footnote{Simulating as if survey response is non-stochastic and not affected by treatment assignments also seems reasonable for two reasons. First, an assumption about the independence of assignment and response would also be required for consistency in the original analysis given by \cite{paluckgreen}. Second, similar to the case of the sharp null, if survey response was independent of assignment but {stochastic}, then variance estimators would be conservativeness relative to the assumption of response being non-random. Again, with this assumption, simulation results give a ``worst-case" assessment of anti-conservative bias.
}

\subsection{Simulation results}

Tables \ref{table.OLSwCov} and \ref{table.OLSnoCov} give simulation results for OLS with and without covariates, respectively, using the data originally reported in \cite{paluckgreen}. Comparing rMSE of variance estimators:

Comparisons:
\begin{itemize}
	\item Generalized Sandwich is less biased and has a smaller SE[$\widehat{\tilde{ \V}}]$ compared to the classical Cluster Robust variance estimators
	\item OC estimators have smaller SE[$\widehat{\tilde{ \V}}]$ compared to GS, while having he same expected value as GS by construction
	\item Only GC$J$ is unbiased for the variance of the OLS coefficient.  It has slightly higher SE[$\widehat{\tilde{ \V}}]$ than the GS, though the rMSE is lower because GS is biased.
\end{itemize}


\setlength{\extrarowheight}{6pt}
\begin{table}[ht]
	\caption{Comparing Variance  Estimators for OLS, No Covariates}
	\centering \scriptsize \addtolength{\tabcolsep}{-2pt}
	\begin{tabular}{rcHcHcHcccccccccccc}
			\cline{3-8} \cline{10-10}  \cline{12-14} \cline{16-18}
		& & \multicolumn{6}{c}{Classical Cluster Robust} 
		& & \makecell[c]{ Gen. \\ Sandwich }
		& & \multicolumn{3}{c}{\makecell[c]{Oblo\v{z}en\`{e} Chleb\`{i}\v{z}ky}}
		& & \multicolumn{3}{c}{\makecell[c]{Guaranteed Conservative}} \\
			\cline{3-8} \cline{10-10}  \cline{12-14} \cline{16-18}
		& & CR0 & CR0 & CR1 & CR1 & CR2 & CR2  & & GS0 & & OC1 & OC$J$.B & OC$J$ & & GC0.B & GC$J$.B & GC$J$ 	
		\\ 
		\hline
		$\E [ \widehat{\tilde{\V}} / \V ] $ &  & 0.802 & 0.429 & 0.864 & 0.462 & 0.932 & 0.818 &  & 0.842 &  & 0.841 & 0.842 & 0.842 &  & 0.858 & 0.860 & 1.000 \\ 
		$\textnormal{SE}[\widehat{\tilde{\V}}/ \V  ]$
		  &  & 0.121 & 0.097 & 0.130 & 0.104 & 0.144 & 0.167 &  & 0.144 &  & 0.123 & 0.123 & 0.123 &  & 0.128 & 0.129 & 0.132 \\ 
		$\textnormal{Bias}[\widehat{\tilde{\V}} / \V ] $ 
		 &  & -0.198 & -0.571 & -0.136 & -0.538 & -0.068 & -0.182 &  & -0.158 &  & -0.159 & -0.158 & -0.158 &  & -0.142 & -0.140 & 0.000 \\ 
		 $\textnormal{rMSE}[\widehat{\tilde{\V}} / \V  ]$ 
		 &  & 0.232 & 0.579 & 0.188 & 0.548 & 0.160 & 0.247 &  & 0.214 &  & 0.201 & 0.200 & 0.200 &  & 0.191 & 0.190 & 0.132 \\ 
		 $\textnormal{CV}[\widehat{\tilde{\V}} ]$ 
		 &  & 0.150 & 0.225 & 0.150 & 0.225 & 0.155 & 0.204 &  & 0.171 &  & 0.146 & 0.147 & 0.146 &  & 0.149 & 0.150 & 0.132 \\ 
		 \hline \makecell[l]{rMSE ratio:} \\ 
		 vs. GS0  &  & 1.086 & 2.712 & 0.882 & 2.566 & 0.747 & 1.155 &  & 1.000 &  &  &  &  &  &  &  &  \\ 
		 vs. OC$J$ &  & 1.158 & 2.892 & 0.941 & 2.737 & 0.797 & 1.232 &  & 1.066 &  & 1.005 & 1.000 & 1.000 &  &  &  &  \\ 
		 vs. GC$J$ &  & 1.759 & 4.392 & 1.428 & 4.156 & 1.210 & 1.870 &  & 1.619 &  & 1.526 & 1.519 & 1.519 &  & 1.451 & 1.442 & 1.000 \\ 
		 \hline \makecell[l]{$\textnormal{CV}$ ratio:} \\ 
		 vs. GS0 &  & 0.880 & 1.317 & 0.880 & 1.317 & 0.906 & 1.191 &  & 1.000 &  &  &  &  &  &  &  &  \\ 
		 vs. OC$J$  &  & 1.027 & 1.537 & 1.027 & 1.537 & 1.057 & 1.389 &  & 1.167 &  & 0.999 & 1.000 & 1.000 &  &  &  &  \\ 
		 vs. GC$J$ &  & 1.141 & 1.707 & 1.141 & 1.707 & 1.174 & 1.543 &  & 1.296 &  & 1.110 & 1.111 & 1.111 &  & 1.132 & 1.134 & 1.000 
		 \\ \hline
	\end{tabular}\label{table.OLSnoCov}
\end{table}

\setlength{\extrarowheight}{6pt}
\begin{table}[ht]
	\caption{Comparing Variance  Estimators for OLS, With Covariates}
	\centering \scriptsize \addtolength{\tabcolsep}{-2pt}
	\begin{tabular}{rcHcHcHcccccccccccc}
		\cline{3-8} \cline{10-10}  \cline{12-14} \cline{16-18}
		& & \multicolumn{6}{c}{Classic Cluster Robust} 
		& & \makecell[c]{ Gen. \\ Sandwich }
		& & \multicolumn{3}{c}{\makecell[c]{Oblo\v{z}en\`{e} Chleb\`{i}\v{z}ky}}
		& & \multicolumn{3}{c}{\makecell[c]{Guaranteed Conservative}} \\
		\cline{3-8} \cline{10-10}  \cline{12-14} \cline{16-18}
		& & CR0 & CR0 & CR1 & CR1 & CR2 & CR2  & & GS0 & & OC1 & OC$J$.B & OC$J$ & & GC0.B & GC$J$.B & GC$J$ 	
		\\ 
		\hline
		$\E [ \widehat{\tilde{\V}} / \V ] $
		&  & 0.795 & 0.431 & 0.856 & 0.464 & 0.941 & 0.819 &  & 0.802 &  & 0.797 & 0.802 & 0.802 &  & 0.853 & 0.861 & 1.000 \\ 
		$\textnormal{SE}[\widehat{\tilde{\V}}/ \V  ]$
		&  & 0.130 & 0.113 & 0.140 & 0.121 & 0.157 & 0.187 &  & 0.131 &  & 0.094 & 0.095 & 0.095 &  & 0.118 & 0.121 & 0.125 \\ 
		$\textnormal{Bias}[\widehat{\tilde{\V}} / \V ] $ 
		&  & -0.205 & -0.569 & -0.144 & -0.536 & -0.059 & -0.181 &  & -0.198 &  & -0.203 & -0.198 & -0.198 &  & -0.147 & -0.139 & 0.000 \\ 
		$\textnormal{rMSE}[\widehat{\tilde{\V}} / \V  ]$ 
		&  & 0.243 & 0.580 & 0.201 & 0.550 & 0.168 & 0.260 &  & 0.237 &  & 0.224 & 0.220 & 0.220 &  & 0.189 & 0.184 & 0.125 \\ 
		$\textnormal{CV}[\widehat{\tilde{\V}} ]$ 
		&  & 0.164 & 0.261 & 0.164 & 0.261 & 0.167 & 0.228 &  & 0.163 &  & 0.118 & 0.119 & 0.118 &  & 0.138 & 0.140 & 0.125 \\ 
		\hline \makecell[l]{rMSE ratio:} \\ 
		vs. GS0  
	&  & 1.024 & 2.447 & 0.847 & 2.318 & 0.708 & 1.095 &  & 1.000 &  &  &  &  &  &  &  &  \\ 
		vs. OC$J$ 
		&  & 1.104 & 2.640 & 0.914 & 2.500 & 0.764 & 1.181 &  & 1.079 &  & 1.019 & 1.000 & 1.000 &  &  &  &  \\ 
		vs. GC$J$ 
		 &  & 1.940 & 4.638 & 1.605 & 4.393 & 1.342 & 2.075 &  & 1.895 &  & 1.790 & 1.758 & 1.757 &  & 1.506 & 1.472 & 1.000 \\ 
		\hline \makecell[l]{$\textnormal{CV}$ ratio:} \\ 
		vs. GS0 
	&  & 1.007 & 1.606 & 1.007 & 1.606 & 1.027 & 1.400 &  & 1.000 &  &  &  &  &  &  &  &  \\ 
		vs. OC$J$
		&  & 1.384 & 2.207 & 1.384 & 2.207 & 1.412 & 1.923 &  & 1.374 &  & 0.998 & 1.003 & 1.000 &  &  &  &  \\ 
		vs. GC$J$ 
		&  & 1.310 & 2.089 & 1.310 & 2.089 & 1.336 & 1.820 &  & 1.300 &  & 0.944 & 0.949 & 0.946 &  & 1.106 & 1.121 & 1.000 \\ 
		\hline
	\end{tabular}\label{table.OLSwCov}
\end{table}

\pagebreak
\pagebreak

\end{document}